\documentclass{aa}  

\usepackage{graphicx}
\usepackage{txfonts}
\usepackage{booktabs}
\usepackage{multirow}
\usepackage{threeparttable}
\usepackage{orcidlink} 
\usepackage{amsmath}
\usepackage[version=4]{mhchem}	
\usepackage{hyperref}
\hypersetup
{
    colorlinks=true, 
    linkcolor=black, 
    citecolor=blue
}
\begin{document}

   \title{Investigating solid-state \ce{CH3OH} formation with chemical modelling}


   \author{\orcidlink{0000-0002-1227-8435} K.-Y. Huang\inst{1}
   \and \orcidlink{0000-0001-9446-6941} E. Méndez-Robayo
          \inst{2*} \thanks{Corresponding author: \email{ejmendezr@unal.edu.co}}
              \and \orcidlink{0000-0001-8504-8844}
     S. Viti\inst{1,3, 4}
          \and \orcidlink{0000-0001-6617-1046} Mario -A. Higuera -G. \inst{2}
          }

   \institute{ Leiden Observatory, Leiden University, PO Box 9513, NL - 2300 RA Leiden, The Netherlands\\
        \and
             Observatorio Astrónomico Nacional, Universidad Nacional de Colombia, Teusaquillo, Bogotá, Bogota, Colombia\\
         \and 
             Transdisciplinary Research Area (TRA Matter) and Argelander Institut f\"{u}r Astronomie, Universit\"{a}t Bonn, Auf dem H\"{u}gel 71, D-53121 Bonn, Germany \\
         \and
             Department of Physics and Astronomy, University College London, Gower Street, London WC1E6BT, UK
             }

   \date{Received ; accepted }

 
  \abstract
   {Recent Monte Carlo simulations and laboratory studies of interstellar ices have proposed an alternative pathway involving the radical-molecule H-atom abstraction reaction in the overall mechanism of methanol (\ce{CH3OH}) formation in dark molecular clouds.}
  {A computational study was conducted to investigate the contribution of the radical-molecule H-atom abstraction route in \ce{CH3OH} formation in interstellar ices, both in non-shocked and shocked environments, and to examine how the physical conditions of the interstellar medium (ISM) affect the overall \ce{CH3OH} synthesis pathway.}
   {A set of chemical models were ran using the gas-grain chemical code UCLCHEM to systematically explore methanol synthesis in various physical scenarios, including non-shock and low- and high-velocity C-shocks.}
   {This work demonstrated for the first time that, under non-shock and shocked-influenced environments, the primary reaction leading to the formation of methanol in the inner layers of interstellar ices is indeed the radical-molecule H-atom abstraction route. 
   However, such route is dependent on the gas kinetic temperature ($T_{k}$), gas volume density ($n_{H_{2}}$), velocity of the C-shock wave ($v_{shock}$), and cosmic ray ionisation rate ($\zeta$). Furthermore, gaseous formaldehyde may trace C-type shocks and serve to differentiate methanol formation pathways in low-velocity C-shocked environments, as its abundance varies more significantly than that of \ce{CH3OH} with the inclusion of the H-atom abstraction reaction in UCLCHEM. The $\frac{X_{\ce{H2CO}}}{X_{\ce{CH3OH}}}$ ratio thus represents a potential diagnostic tool for this purpose.
   }
   {}


   \maketitle
%

\section{Introduction}

Interstellar methanol is considered an important building block of complex organic molecules (COMs) through surface and gas phase reactions, such as \ce{CH2OH} radical, \ce{CH3O} radical, \ce{CH3CH2OH} (ethanol), \ce{CH3OCH3} (dimethyl ether), \ce{CH3CHO} (methyl formate), \ce{(CH2OH)2} (ethylene glycol) and other species \citep{charnley1992, fuchs2009,cecchi2018}. \cite{mathew2022} discussed how methanol could be the starting material for olefins, carbonyl compounds, amines, amino acids, peptides, polypeptide chains and complex life molecules such as RNA. Hence, it is crucial to investigate and understand the pathways leading to interstellar methanol formation.

The existing framework of interstellar methanol formation involves a vast network of interrelated reactions. Among these, the two formation pathways of most interest are the pure gas phase and the solid-state schemes. 

In the gas phase, protonated methanol  (\ce{CH3OH2+}) can be produced by the radiative association reaction, which is primarily activated by cosmic rays or stellar radiation (Rxn. \eqref{eq: first step gas}) \citep{cecchi2018,luca2002}. 

\begin{equation}
\label{eq: first step gas}
    \ce{CH3+ + H2O -> [CH3OH2+]^{\ast} -> CH3OH2+} + h\nu,
\end{equation}

In this radiative association reaction, the protonated methanol is formed in a highly exothermic reaction, \ce{\Delta H
= -2.9} eV \citep{luca2002}, where the collision complex,  \ce{[CH3OH2+]^{$\ast$}}, is formed in an energetic excited state and then stabilized by emitting a photon. \cite{geppert2006} argue that in more complicated systems, the radiative lifetime of the excited complex can be of the same order of magnitude as the dissociative lifetime. 
Therefore, both processes  may compete in determining the stability of the excited complex and, hence, affecting the overall production of protonated methanol.
The protonated methanol then dissociatively recombines with electrons to form methanol (Rxn. \eqref{eq: second step gas}) \citep{cecchi2018, geppert2006, yamamoto2017}:

\begin{equation}
\label{eq: second step gas}
    \ce{CH3OH2+ + e -> [CH3OH2]^{\ast} -> CH3OH + H}.
\end{equation}

This last reaction, at \ce{\Delta H = -5.7}eV  is one of the few dissociative recombination pathways possible in the gas-phase ISM due to the low kinetic energies prevailing there \citep{geppert2006}. \cite{wirstrom2011, geppert2005}, however, had established that the dissociative recombination step, Rxn.\eqref{eq: second step gas}, cannot be efficient enough for a pure gas phase route to alone explain the observed \ce{CH3OH} abundances. 

On the other hand, solid-state methanol is an abundant  constituent of ice-type mantles \citep{mcclure2023}.  In dense and cold regions of molecular clouds, \ce{CO} molecules present in the gas phase freeze out and form an apolar coating on top of icy-dusty grains \citep{cecchi2018, chuang2016, chuang2018, linnartz2015, santos2022}. Then, methanol can be efficiently formed through the continuous hydrogenation of \ce{CO} in the solid phase, with formaldehyde (\ce{H2CO}) as one of the intermediate products (see \ce{CO} hydrogenation mechanism in Table \ref{tab:rxn-rates}).

The \ce{CO} hydrogenation, which is a Langmuir-Hinshelwood (L-H) mechanism, has been regarded as the primary route to producing methanol in dark molecular clouds. In addition, this process has been shown to effectively reproduce methanol abundances observed in translucent clouds, while pure gas phase models fail by four orders of magnitude \citep{turner1998}. However, recent laboratory results and Monte Carlo simulations suggest that the radical-molecule H-atom abstraction route (Rxn. \eqref{eq: second grain step} \citep{santos2022, alvarez2018, simons2020}), also a L-H mechanism, as the dominating (70-90\%) final step to form \ce{CH3OH} in dark molecular clouds, replacing hydrogenation of \ce{CH3O} in the overall \ce{CH3OH} formation mechanism \citep{santos2022, alvarez2018, simons2020}.  
\cite{santos2022} suggests that the total contribution of each pathway is determined by the availability of \ce{H}, \ce{CH3O}, and \ce{H2CO} in the ice. 

\begin{equation} 
 \label{eq: second grain step}
  \ce{CH3O + H2CO -> CH3OH + HCO}.
\end{equation}

In addition, under \ce{H2O}-rich interstellar ice environments, \cite{qasim2018} studied the \ce{CH3OH} formation in ice analogues and  found that methanol formation is possible through the sequential surface reaction chain: \ce{CH4 + OH -> CH3 + H2O} and \ce{CH3 + OH -> CH3OH}. In addition, \cite{molpeceres2021} suggest the possibility that \ce{CH3OH} could form via \ce{C + H2O} in water-rich ices. However, \cite{qasim2018} conclude that the \ce{CO + H} reaction pathway is approximately 20 times more effective in producing \ce{CH3OH} at 10 K compared to the \ce{CH4 + OH} route. Furthermore,\cite{2025jimenezserra} model the ice chemistry in the Chamaeleon I molecular cloud using a set of state-of-the-art astrochemical codes and found out that \ce{CH3OH} ice formation occurs predominantly (>99\%) via \ce{CO} hydrogenation, while the contribution of reactions \ce{CH3 + OH} and \ce{C + H2O} is negligible.

The present work aims to study the main interstellar methanol formation route under non-shocked and shocked scenarios using the open source gas-grain chemical code UCLCHEM v.3.1 \footnote{https://uclchem.github.io/docs/} \citep{holdship2017}. This study explores  the influence of the radical-molecule H-atom abstraction route in methanol formation under shock-influenced environments, such as galactic protostellar outflows. In Section 2, we describe the chemical models performed. In Section 3, we assess the importance of  gas-phase versus solid-phase  methanol formation mechanisms  in the framework of   observations of outflows. In Section 4, we analyse the impact of the radical-molecule H-atom abstraction route on \ce{HCO}, \ce{H2CO}, and \ce{CH3O}. We briefly summarize our findings in Section 5.

\section{Chemical modelling of methanol}

We use the time dependent open source gas-grain chemical model UCLCHEM \citep{holdship2017} to model methanol formation in star forming regions.

Several physics modules are available in UCLCHEM. The so called "Hot Core" module allows us to model the envelopes surrounding a forming protostar.  In this model, the temperature increases following the time and radially dependent temperature profiles described by \cite{viti2004}, and adapted in \cite{awad2010}. The "shock" modules (C- and J-) are instead used to model shocked regions. For this module,  UCLCHEM adopts a parametrised C-shock and J-shock from \cite{jimenez2008} and \cite{james2020} respectively, to describe the changes in gas properties over time in the presence of shocks. For this study, we just use the C-shock module. 

For the C-shock models, we run UCLCHEM for three epochs, according to the epoch of shock propagation. The pre-shock epoch corresponds to the time in which the initial gas kinetic temperature, $T_{k}\: (t)$ remains low and is the same as $T_{k, init}$. The beginning of the shock is determined by the rise of $T_{k}\: (t)$ due to shock heating, and the post-shock epoch is determined by its decrease until it reaches the initial cold temperature again. For gas phase chemistry, we use the UMIST12 database \citep{mcelroy2013}. The surface network is supplied in the UCLCHEM github and includes the standard hydrogenation route to methanol from \ce{CO}. We shall call this network the "standard" network. 

To implement the radical-molecule route (Rxn.\eqref{eq: second grain step}) in the UCLCHEM network, we introduce the reaction list given in Table \ref{tab:rxn-rates}. 
In the added network, we use the branching ratios proposed by \cite{simons2020} and the barrier energies reported in KIDA DATABASE \citep{wakelam2012}  and \cite{alvarez2018} for implementing the radical-molecule route into the UCLCHEM network. We shall call this augmented network the "current" network. 

\begin{table*}[t]
\centering
\caption{Methanol formation reaction rate constants.}
\label{tab:rxn-rates}
\begin{tabular}{@{}ccccccc@{}}
\toprule
\multirow{2}{*}{\textbf{Reactions}} & \multicolumn{3}{c}{\textbf{Current}} & \multicolumn{3}{c}{\textbf{Default}} \\ \cmidrule(lr){2-4} \cmidrule(l){5-7}
 & $\alpha$ & $\beta$ & $\gamma$ & $\alpha$ & $\beta$ & $\gamma$ \\ \midrule
\ce{CH3+ + H2O -> CH3OH2+}+ h$\nu$ & 2E-12$^{a}$ & 0.0 & 0.0 & 2E-12$^{a}$ & 0.0 & 0.0 \\
\ce{CH3OH2+ + e -> CH3OH + H} & 2.67E-8$^{b}$ & -0.59$^{b}$ & 0.0 & 2.67E-8$^{b}$ & -0.59$^{b}$ & 0.0 \\ \midrule
\ce{H + CO -> HCO} & 1.0$^{e}$ & 0.0 & 2500$^{d}$ & 1.0$^{d}$ & 0.0 & 2500$^{d}$ \\
\ce{H + HCO -> H2CO} & 0.417$^{e}$ & 0.0 & 0.0 & 0.5 & 0.0 & 0.0 \\
\ce{H + HCO -> CO + H2} & 0.583$^{e}$ & 0.0 & 0.0 & 0.5 & 0.0 & 0.0 \\
\ce{H + H2CO -> CH2OH} & 7.27E-5$^{e}$ & 0.0 & 5400$^{c}$ & 0.33$^{c}$ & 0.0 & 5400$^{c}$ \\
\ce{H + H2CO -> CH3O} & 0.33$^{e}$ & 0.0 & 2200$^{c}$ & 0.33$^{c}$ & 0.0 & 2200$^{c}$ \\
\ce{H + H2CO -> HCO + H2} & 0.33$^{e}$ & 0.0 & 1740$^{c}$ & 0.33$^{c}$ & 0.0 & 1740$^{c}$ \\
\ce{H + CH2OH -> CH3OH} & 0.6$^{e}$ & 0.0 & 0.0 & 1.0 & 0.0 & 0.0 \\
\ce{H + CH3O -> CH3OH} & 0.3$^{e}$ & 0.0 & 0.0 & 1.0 & 0.0 & 0.0 \\
\ce{CH3O + H2CO -> CH3OH + HCO} & 0.4$^{e}$& 0.0 & 2670$^{f}$ & - & - & - \\ \bottomrule
\end{tabular}
\begin{tablenotes}
\footnotesize
\item $^{a}$ \cite{luca2002}, $^{b}$ \cite{geppert2006}, $^{c}$ UCLCHEM basic network, $^{d}$ \cite{wakelam2012}, $^{e}$ \cite{simons2020}, $^{f}$ \cite{alvarez2018}.
\end{tablenotes}
\end{table*}

Table \ref{tab:conditions} summarizes the physical conditions used to model the non-shock and shock chemistry in various physical scenarios. We ran 192 models in total, taking into account both default and current networks. The UCLCHEM modelling consists of two stages ordered in time sequence: in stage 1 diffuse gas (initial $n_{H_{2}\:(init)}=10$ $cm^{-3}$) undergoes free-fall collapse until a density set by the user (See Table 2). During this stage, the gas evolves from elemental atomic to the composition of a dense cloud.  In the second stage, we then evolve the chemistry under two scenarios: C-type shocks and "Hot Cores", with the latter representing  dense gas around protostars that has not been shocked.

\begin{table}[h]
\centering
\caption{Range of Values for Stage-2 Variables}
\label{tab:conditions}
\resizebox{\columnwidth}{!}{
\begin{tabular}{@{}cc@{}}
\toprule
\textbf{Variable}                             & \textbf{Value}                \\ \midrule
Hot Core gas temperature,  $T_{HC}$$ \: [K]$ & 50, 100, 200                  \\
Gas density, $n_{H_{2}\:(init)} \: [cm^{-3}]$ & $10^{3}, 10^{4}, 10^{5}, 10^{6}$ \\
CRIR, $\zeta \: [\zeta_{0}]$                  & 1, 10, 100                    \\
Extinction, $A_{v} \: [mag]$                  & 10                            \\
C-shock velocity, $\varv_{shock}\: [km/s]$    & 10, 20, 30, 40, 50                        \\
Physical model                                & Shock or non-shocked scenario \\
Network                                       & Current or default            \\ \bottomrule
\end{tabular}}
\end{table}

During the first stage, the gas kinetic temperature of $T_{k, init}=10$ K and the  cosmic ray ionization rate (CRIR) for molecular hydrogen is standard at  $\zeta_{0}= 5\times10^{-17}$ s$^{-1}$.  The cloud was allowed to evolve for $\sim 10^{7}$ years.

For the C-shock scenario, we varied the CRIR from $\zeta=\zeta_{0}$ to $\zeta=100\zeta_{0}$, and the shock velocity ($\varv_{shock}$) from $10$ to $50$ km/s. The chosen range for $\varv_{shock}$  allows us to distinguish between a low-velocity shock ($10$ km/s as a representative value) and a high-velocity shock ($50$ km/s as a representative value). In high-velocity shocks, sputtering of \ce{Si} from the grain-core occurs. Meanwhile, the non-shock model was simulated by heating up the cloud (composed of one gas parcel) to a final gas kinetic temperature, defined as the Hot Core temperature $T_{HC}$, of $T_{HC}=$50, 100, and 200 K, by considering a $10 M\odot$ embedded protostar \citep{viti2004}. For both shock and non-shock scenarios, the cloud was allowed to evolve for $10^{7}$ years for the second stage as well.


\section{Interstellar \ce{CH3OH} Formation}\label{section:formation}

This Section presents a detailed analysis of the behaviour and formation pathways of solid-state methanol, both on the surface and in the inner layers (bulk) of the icy-dusty grain. The aim of this analysis is to verify whether the  implemented radical-molecule H-atom abstraction reaction does become the dominant final step in forming solid-state \ce{CH3OH} as stated in \cite{simons2020} and \cite{santos2022}. The term 'ice phase' will refer to the combined contribution of the grain surface and bulk. 

\begin{figure*}[t]
            \sidecaption
            \includegraphics[width=12cm]{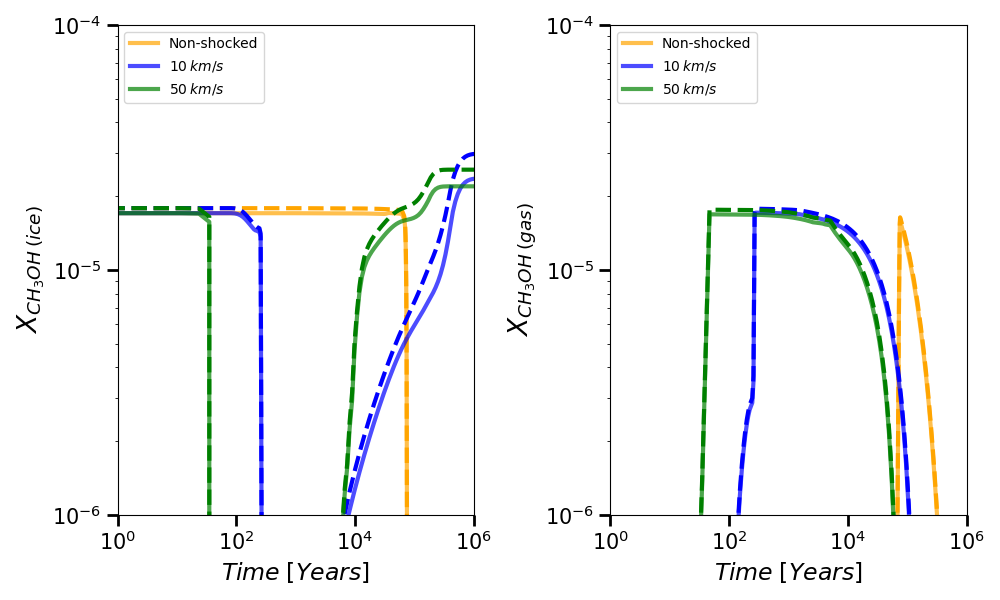}
            \caption{Chemical abundances of ice-phase methanol (left panel) and gas-phase methanol (right panel) as a function of time in non-shocked, slow C-shock (10 km/s) and high C-shock (50 km/s) models with a pre-shocked gas density volume of $n_{H_{2\:(init)}}=10^{4}\: cm^{-3}$ and $\zeta = \zeta_{0}$. The dashed and solid lines correspond to the default and current schemes, respectively.} \label{fig:ice_1}
        \end{figure*}

In non-shock scenarios, the abundance of methanol in the ice remains stable throughout the model until $10^{5}$ yrs. Any following decrease as $T_{k}\:(t)$ reaches up to 200 K is attributed to the incidence of cosmic rays or thermal desorption. However, when a C-shock is present regardless of its velocity, there is a significant decrease in the abundance of the ice-phase as the shock propagates (see left panel in Fig \ref{fig:ice_1}.). The destruction of ice layers during the propagation of the C-shock wave results in the release of methanol from the solid to the gas phase, leading to a significant decrease in solid-state methanol abundance. As the C-shock wave dissipates, ice mantles begin to reform, and methanol gradually re-accumulates in the solid phase as new layers grow on the grain surfaces, which enhances the abundance of solid-state methanol (see right panel in Fig \ref{fig:ice_1}.). The behaviour of \ce{CH3OH} in the solid-state changes when a C-shock propagates, so it is pertinent to analyse the reactions that form \ce{CH3OH}, both on the surface and in the bulk of the grain, and whether the overall \ce{CH3OH} synthesis pathway vary according to the gas kinetic temperature ($T_{k}$), gas volume density ($n_{H_{2}}$), velocity of the C-shock wave ($v_{shock}$), and cosmic ray ionisation rate ($\zeta$).

The following subsections analyse each \ce{CH3OH}-state and the differences between the most important reactions at different time steps. 

\subsection{Grain-surface phase}

 \begin{figure*}[t]
            \sidecaption
            \includegraphics[width=12cm]{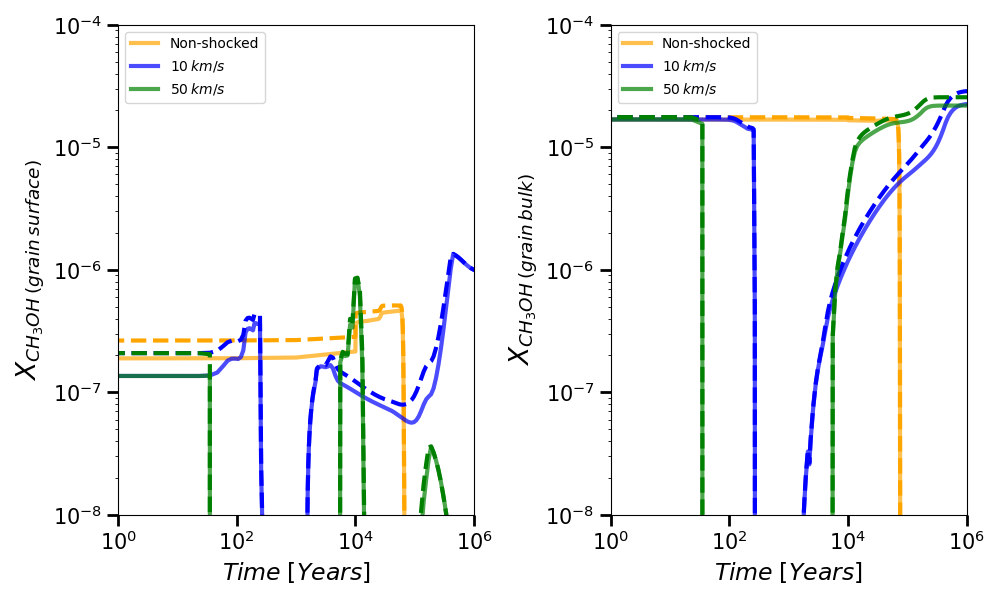}
            \caption{Chemical abundances of surface-phase methanol (left panel) and bulk-phase methanol (right panel) as a function of time in non-shocked, slow C-shock (10 km/s) and high C-shock (50 km/s) models with a pre-shocked gas density volume of $n_{H_{2\:(init)}}=10^{4}\: cm^{-3}$ and $\zeta = \zeta_{0}$. The dashed and solid lines correspond to the default and current schemes, respectively.} \label{fig:ice_2}
        \end{figure*}

In non-shock models, surface methanol (see left panel in Fig \ref{fig:ice_2}.) is still primarily formed through the hydrogenation of \ce{CH3O} (Rxn.\eqref{eq:h3co hydro solid}) and \ce{CH2OH} (Rxn.\eqref{eq:ch2oh hydro solid}).  The prevalence of hydrogenations may stem from the fact that these \ce{H}-atom addition routes are promoted with a higher \ce{H}-atom flux to the surface, allowing the accreted \ce{H}-atoms to diffuse and react with both \ce{CH3O} and \ce{CH2OH} species. Methanol synthesis through \ce{CH3O} or \ce{CH2OH} hydrogenation is a process that has no barriers \citep{santos2022}, making it ideal for enhancing methanol production at such low temperatures. Once $T_{k}\:(t)$ = $T_{HC}$,  the condensations of gaseous \ce{CH3OH} and \ce{CH3OH2+} (Rxn.\eqref{eq:condens solid}) ascribe to the surface methanol formation. 

\begin{align}
    &\ce{CH3O_{(surf)} + H_{(surf)} -> CH3OH_{(surf)}}, \label{eq:h3co hydro solid}\\
    &\ce{CH2OH_{(surf)} + H_{(surf)} -> CH3OH_{(surf)}}, \label{eq:ch2oh hydro solid}\\
    &\ce{CH3OH2+_{(g)} -> CH3OH_{(surf)} + H_{(g)}}. \label{eq:condens solid}
\end{align}

However, the formation of surface methanol during a shock involves an additional series of reactions. The H-atom addition reactions  are replaced by the H-abstraction route (Rxn.\eqref{eq:rxn abstract solid}\citep{simons2020}) as $T_{k}\:(t)$ rises in low-velocity C-shocks, while bulk-to-surface transfer dominates at the maximum $T_{k}\:(t)$.

\begin{align}
    &\ce{CH3O_{(surf)} + H2CO_{(surf)} -> CH3OH_{(surf)} + H_{(surf)}}. \label{eq:rxn abstract solid}
\end{align}

The main difference between the default and current networks is the prevalence of the radical-molecule \ce{H}-atom abstraction route as $T_{k}\:(t)$  increases in low-velocity shocks. In the current network, methoxy radical hydrogenation dominates as the final step in the overall \ce{CO} hydrogenation mechanism on the grain surface. However, as the temperature rises due to shock heating, the abstraction route replaces radical \ce{CH3O} hydrogenation as the final step in \ce{CH3OH} formation. Hence the methanol yield from both networks is dependent on the surface temperature.  \cite{santos2022} and \cite{simons2020} show that there is a competition between the increase in the diffusion of \ce{H} atoms and the decrease in their residence time on the ice as a function of temperature. At low temperatures, \ce{CH3O} surface radicals undergo hydrogenation. However, as the temperature increases to above 20-30 K, the residence time of \ce{H} atoms on the surface decreases substantially. At the same time, radicals become mobile, allowing the abstraction reaction to become competitive with \ce{CH3O} radical hydrogenation. 

This finding seems to be in contradictions with the results from   \cite{simons2020} and \cite{santos2022}. The former finds that the \ce{CH3O + H} pathway becomes significant only at higher temperatures (although their study only covers temperatures up to 20K), and when the surface coverage of \ce{H2} is low.  The latter reports experimental evidence suggesting that the abstraction mechanism may already dominate at lower temperatures (10–16 K). However, their study does not distinguish between surface and bulk ice chemistry in the same way as the present model. This disagreement may stem from various reasons: (i) incomplete surface networks in our model; (ii) our different treatment of surface of the grains where we distinguish between bulk and surface; (iii) the different physical conditions such as density and temperature ranges. The latter in particular increases extremely fast in our shocks models and the high temperature lasts for a very short time, before returning to 10 K.

\subsection{Grain-bulk phase}

As $T_{k}\:(t)$ increases, in both non-shock and shocked models (see right panel in Fig \ref{fig:ice_2}.), methanol is primarily formed through the hydrogenation of \ce{CH3O} (Rxn.\eqref{eq:h3co hydro bulk}\citep{simons2020}, the radical-molecule route (Rxn.\eqref{eq:abstract bulk}\citep{simons2020}, and surface-to-inner layer transfer. Transfer from surface to inner layers and the hydrogenation of \ce{CH3O} dominate until $T_{k}\:(t)$$\sim$ 70 K at $n_{H_{2\:(init)}}=10^{3}\:cm^{-3}$, after which the H-abstraction route takes over. At $n_{H_{2\:(init)}}\geq10^{4}\:cm^{-3}$, the primary methanol formation routes are the H-abstraction reaction and surface-to-inner layer transfer.

\begin{align}
    &\ce{CH3O_{(bulk)} + H_{(bulk)} -> CH3OH_{(bulk)}}, \label{eq:h3co hydro bulk}\\
    &\ce{CH3O_{(bulk)} + H2CO_{(bulk)} -> CH3OH_{(bulk)} + HCO_{(bulk)}}. \label{eq:abstract bulk}
\end{align}

The radical-molecule H-atom abstraction route dominates the formation of methanol in the bulk, with the gas density being the most influential factor. High $n_{H_{2}}$ favours the formation of \ce{H2} and \ce{OH} on the grain surface (Rxns.\eqref{eq:h2surf}\citep{cazaux2004}, \eqref{eq:cosurf}\citep{simons2020}, and \eqref{eq:osurf}\citep{wakelam2012}, which accumulates in the bulk as new surface layers form. Due to its high reactivity, the \ce{H} atom reacts with \ce{CO} and \ce{O} to primarily form \ce{H2O} and \ce{H2CO} in the bulk grain (Rxns.\eqref{eq:cobulk}\citep{simons2020}, \eqref{eq:cobulk2}\citep{simons2020}, and \eqref{eq:obulk}\citep{wakelam2012}). The \ce{H} atoms must diffuse in order to react with the minor ice component, \ce{CH3O} radicals. Therefore, the dominance of the radical-molecule route is likely due to the greater availability of \ce{H2CO} in the bulk, compared to \ce{H} atoms, for reacting with \ce{CH3O}, as stated in \cite{santos2022}.

\begin{align}
 &\ce{H_{(surf)} + H_{(surf)} -> H2_{(surf)}},\label{eq:h2surf}\\
 &\ce{CO_{(surf)} ->[H_{(surf)}] HCO_{(surf)}  ->[H_{(surf)}] CO_{(surf)} + H2_{(surf)}},\label{eq:cosurf}\\
 &\ce{H_{(surf)} + O_{(surf)} -> OH _{(surf)}}.\label{eq:osurf}
\end{align}

\begin{align}
 &\ce{CO_{(bulk)} ->[H_{(bulk)}] HCO_{(bulk)} ->[H_{(bulk)}] CO_{(bulk)} + H2_{(bulk)}},\label{eq:cobulk}\\
 &\ce{CO_{(bulk)} ->[H_{(bulk)}] HCO_{(bulk)} ->[H_{(bulk)}] H2CO_{(bulk)}},\label{eq:cobulk2}\\
 &\ce{O_{(bulk)} ->[H_{(bulk)}] OH_{(bulk)}   ->[H_{(bulk)}] H2O_{(bulk)}}.\label{eq:obulk}
\end{align}

\subsection{Ice phase}

In our model, the hydrogen abstraction reaction is more relevant in the bulk phase of the ice mantle, whereas \cite{simons2020, santos2022, 2025jimenezserra} present their results in terms of the grain-surface mantle or ice-grain without explicitly distinguishing between the surface and the interior, as we do by defining the former as the outermost layer and the latter as everything beneath the surface. Since we consider the grain to be largely composed of bulk material, we can extend our observation to the ice phase in general.

\cite{simons2020} performed a series of simulated co-depositions of \ce{CO}+\ce{H} and \ce{H2CO}+\ce{H} considering different H-atom binding energies (250 K, 420 K, and 670 K). In their simulations, the authors found that the radical-molecule route was the primary pathway for methanol formation. They emphasized the dependence of \ce{CH3OH} formation on the \ce{H}:\ce{CO} ratio. They also pointed out that the dominance of the H-abstraction reaction was because when two \ce{H2CO} molecules were produced nearby, only one \ce{H} atom was required for hydrogenation to convert \ce{CH3O} into \ce{CH3OH} and \ce{HCO}. Subsequently, these species underwent rapid hydrogenation once more.

On the other hand, \cite{santos2022} conducted experiments under ultra-high vacuum conditions and astronomically relevant temperatures. They employed \ce{H}:\ce{H2CO} (or \ce{D2CO}) flux ratios of \ce{10}:\ce{1} and \ce{30}:\ce{1}. Similar to \cite{simons2020}, \cite{santos2022} observed that the radical-molecule route was the predominant pathway for methanol formation in their experiments. They specifically tested a \ce{H}/\ce{H2CO}(\ce{D2CO}) ratio of 30 at 10 K and concluded that the \ce{CHD2OH}/\ce{CH3OH} ratios were negligibly different, considering the detection error. Consequently, \cite{santos2022} suggested that a higher hydrogen flux only slightly affected the contribution of the radical-molecule H-atom abstraction route, even if a potentially higher \ce{H} flux might favor the atom-addition route, and still governed the formation of \ce{CH3OH}. Our results confirm the last statement for the range of $n_{H_{2\:(init)}}=10^{4}-10^{6}\:cm^{-3}$. In this range, the H-abstraction route dominates and produces alike methanol abundances. Therefore, the product yield is limited by the abundance of formaldehyde rather than the abundance of hydrogen atoms, which are in excess.  These results also support the findings of \cite{santos2022}, where the contribution from each route is determined by the availability of hydrogen, formaldehyde and methoxide radical in the ice.  In an H-abstraction reaction, the probability of the two reactants coming into contact is higher compared to a barrier-less hydrogenation reaction.

\cite{santos2022} reports that the abstraction route accounts for approximately 80\% of the contributions in the 10-16 K interval, which is consistent with the findings of \cite{simons2020}. In both studies, \ce{CH3OH} formation is independent of temperature in the 10-16 K range. Moreover, our results indicate that the H-abstraction reaction is the dominant final step in the \ce{CO} hydrogenation to form \ce{CH3OH}, both at low gas kinetic temperatures, such as those found in molecular clouds, and at high gas kinetic temperatures, such as those achieved by shock heating, under $n_{H_{2\:(init)}}=10^{4}-10^{6}\:cm^{-3}$. Quantum tunnelling is activated at low temperatures, but as the temperature rises, it probably becomes also thermally activated. This last scenario is most commonly observed under low $n_{H_{2\:(init)}}$, where elevated temperatures may promote the radical-molecule route.

We  note that in \cite{2025jimenezserra}, multiple models were ran utilizing various astrochemical codes to analyse the ice abundances recorded by the JWST during the Ice Age ERS program \citep{mcclure2023}, focusing on the heavily obscured background stars NIR38 and J110621 in the line of sight of the dense molecular cloud Chameleon I. In the stochastic Kinetic Monte Carlo (KMC) simulations \citep{2007cuppen} that examine the H-abstraction pathway for the synthesis of \ce{CH3OH}, it was determined that methanol is primarily produced via the sequential hydrogenation of \ce{CO}. They determined that approximately 60\% is primarily derived from the hydrogenation of \ce{CH2OH}, which they did not classify as a formation reaction since \ce{CH2OH} is exclusively generated through H-abstraction of methanol. Around 30\% has been produced via the hydrogenation of \ce{CH3O}, while the radical-neutral reaction \ce{CH3O + H2CO -> CH3OH + HCO} contributes only about 10\%. The discrepancies may arise from the varying grid of conditions employed in both studies (e.g., reduced extinction usage in this research). 

In terms of CRIR, cosmic ray-induced photons can dissociate \ce{H2CO} and \ce{CH3OH}, producing functional group radicals such as \ce{CH3} and \ce{CH3O} on or within the ice.  Our results also suggest that high CRIR ($\zeta=100\zeta_{0}$) decreases the abundance of methanol in all phases. However, this rate affects both the radical-molecule route and other possible pathways, as the energy from cosmic rays can break apart precursors molecules involved in the overall \ce{CO} hydrogenation mechanism before they can react.

\begin{figure}[h]
            \centering
            \resizebox{\columnwidth}{!}{
            \includegraphics[width=350pt]{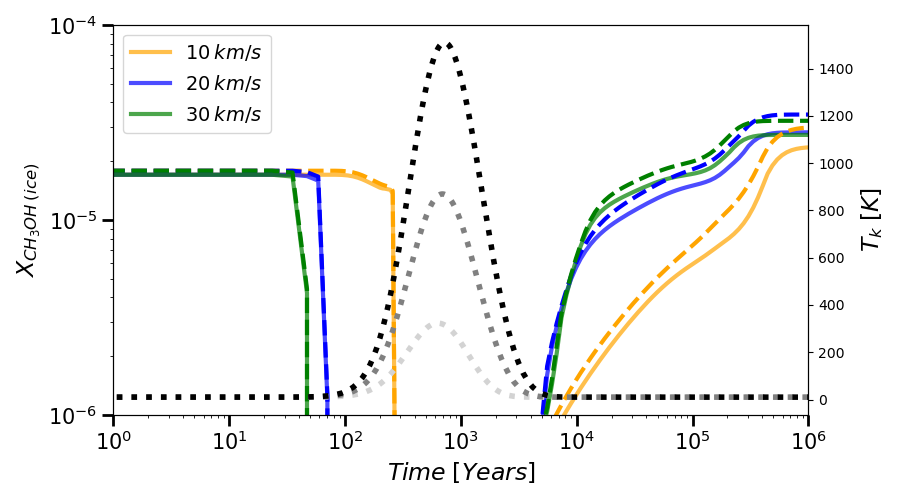}}
            \caption{Chemical abundances of ice-phase methanol as a function of time in slow C-shock (10-20 km/s) and moderate-high C-shock (30 km/s) models with a pre-shocked gas density volume of $n_{H_{2\:(init)}}=10^{4}\: cm^{-3}$ and $\zeta = \zeta_{0}$. The dashed and solid lines correspond to the default and current schemes, respectively. The evolution of gas temperature, $T_{k}\:(t)$, is depicted by the gray dotted lines.} \label{fig:grain_1}
        \end{figure}

Figure \ref{fig:grain_1} show that in low velocity C-shocks, ice-methanol is more abundant when using the default network compared to the current one. The main difference between the two networks is that the H-abstraction reaction is the most important pathway for methanol formation in the bulk, while the \ce{CH2OH} hydrogenation in the grain surface has become more significant due to changes in branching ratios, especially at higher $T_{k}\:(t)$. With the current network, we use branching ratios from \cite{simons2020}. 
These rates were based on quantum chemical calculations combined with a microscopic kinetic Monte Carlo simulation, and  yield a lower but proper amount of methanol. It should be noted that the predicted difference in bulk methanol abundance between networks is minimal ($\sim 1\times10^{-5}$,$\frac{X_{default}}{X_{current}}\leq1.5$) because our network does not include other reactions (i.e.,  \ce{HCO} + \ce{HCO} reactions \citep{simons2020}) that could also alter the molecular composition of the grain mantle. Therefore, we encourage the extension of the chemical reaction network for the sake of completeness in future works. 

\subsection{Comparison with Observations}

It is of interest to ascertain whether modifying the solid-state reaction network would have a direct effect on the formation of \ce{CH3OH_{(g)}} by introducing different abundances in the gas-phase. As mentioned in \cite{Huang2025} and \cite{huangviti2023}, gas-phase methanol is a potential tool to trace C-type shocks, as shocks can substantially enhance the gas-phase abundance of \ce{CH3OH}. Therefore, we aim to analyse whether the changes in the solid phase formation pathway alter the observable abundance of gas phase methanol as the shock sputters the grain as it propagates.

Gas-phase methanol is mainly due to \ce{CH3O} hydrogenation and subsequent chemical desorption (Rxn.\eqref{eq:hydro1}\citep{simons2020, minissale2016}), with an important contribution from \ce{CH2OH} hydrogenation (Rxn.\eqref{eq:hydro2}\citep{simons2020, minissale2016}) in the models at $n_{H_{2\:(init)}}=10^{3}\: cm^{-3}$. As the gas kinetic temperature increases due to shock heating, the hydrogenation of \ce{CH3O} is generally replaced by  multiple reactions such as ion-neutral Rxn. \eqref{eq: gasrxn_h3+}\citep{sung1992}, protonated methyl alcohol (\ce{CH3OH2+}) recombination in Rxn. \eqref{eq: recombination ch3oh2+}\citep{geppert2006}, and the ion-neutral Rxn.\eqref{eq: gasrxn_nh3}\citep{wakelam2012}. 

\begin{align}
& \ce{H_{(s)} + CH3O_{(s)} -> CH3OH_{(g)}}, \label{eq:hydro1}\\
& \ce{H_{(s)} + CH2OH_{(s)} -> CH3OH_{(g)}}. \label{eq:hydro2}\\
&\ce{H3+ + CH3CHO -> CH3+ + CH3OH}, \label{eq: gasrxn_h3+}\\
&\ce{CH3OH2+ + e- -> CH3OH + H}, \label{eq: recombination ch3oh2+}\\
&\ce{CH3OH2+ + NH3 -> CH3OH + NH4+}.  \label{eq: gasrxn_nh3}
\end{align}

However, the radical-molecule route, Rxn.\eqref{eq:nrxn_sg} \citep{simons2020,minissale2016, santos2022}, has a impact on gaseous methanol production only at low-velocity C-shocks with high initial gas volume densities ($n_{H_{2\:(init)}}=10^{5}-10^{6}\: cm^{-3}$). 

\begin{align}
& \ce{H2CO_{(s)} + CH3O_{(s)} -> CH3OH_{(g)} + H_{(g)}}. \label{eq:nrxn_sg}
\end{align}

Therefore, it is pertinent to contrast our UCLCHEM models with galactic protostellar outflow observations made by \cite{holdship2019} to ascertain if it is possible to observationally distinguish the most favourable pathways for \ce{CH3OH_{(s)}} formation under shock-induced environments.

Class 0 young stellar objects, like NGC 1333-IRAS2A and NGC 1333-IRAS4A are associated with outflows. These outflows create shocks upon impacting the surrounding medium \citep{dishoek1998, holdship2019}. As \cite{holdship2019} observed, these shock regions showcase a diverse chemistry, with gas-phase methanol experiencing notable increases in abundance. In order to model the physical structure trace by \ce{CH3OH} in the protostellar outflow sources - IRAS2A-R, IRAS2A-B, IRAS4A-R, and IRAS4A-B - we used the inferred gas properties ($n_{H_{2}\:(init)}$) from \cite{holdship2019}.  Our grid of models include these inferred gas properties to study the influence of formation pathways (processing of interstellar ices and pure gas phase mechanisms) on the observed interstellar methanol abundances. 

\begin{figure}[h]
            \centering
            \resizebox{\columnwidth}{!}{
            \includegraphics[width=350pt]{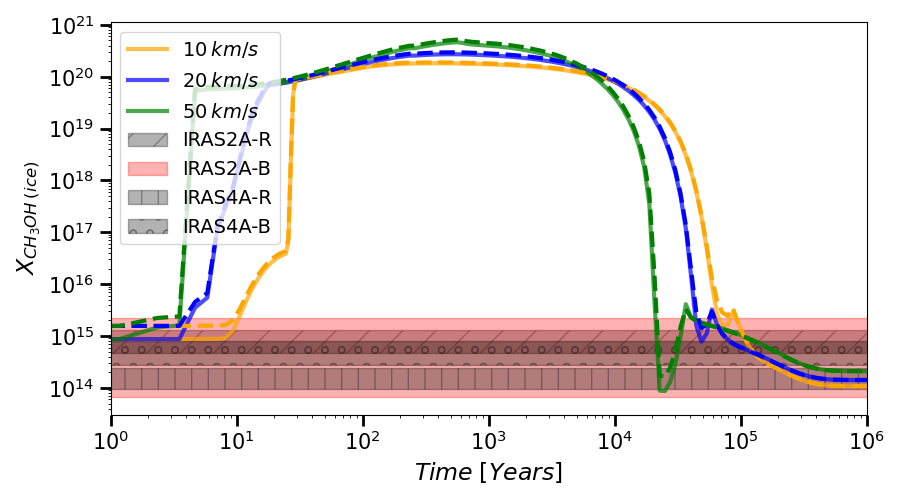}}
            \caption{Column densities of gaseous \ce{CH3OH} as a function of time for slow C-shock (10-20 km/s) and high C-shock (50 km/s) models with a pre-shocked gas density volume of $n_{H_{2\:(init)}}=10^{5}\: cm^{-3}$ and $\zeta = \zeta_{0}$. The dashed and solid lines correspond to the default and current schemes, respectively. The hatches represent the range in \ce{CH3OH} column density reported by \cite{holdship2019} for IRAS2A-R, IRAS2A-B, IRAS4A-R, and IRAS4A-B.} \label{fig:gas_ch3oh_iras}
        \end{figure}

To effectively compare the models with observations, we focus on models with an abundance exceeding $X_{\ce{CH3OH}}=10^{-7}$, and where the ratio between the standard and the current network, $\frac{X_{default}}{X_{current}}\geq2$. For gas-phase methanol, just the model with $n_{H_{2\:(init)}}=10^{5}$, $\zeta=\zeta_{0}$, and $\varv_{shock}=20 km/s$ fulfills both characteristics at the same time.  Figure \ref{fig:gas_ch3oh_iras} illustrates a comparison of gas-phase methanol column densities over time using 10-50 km/s models  with the range of methanol column densities reported by \cite{holdship2019} for IRAS2A-R, IRAS2A-B, IRAS4A-R, and IRAS4A-B. In order to compare our models directly with the observations, we derived column densities from the modelled fractional abundances by using the on-the-spot approximation:
\begin{equation}
N(X) = X \times A_V \times B(H_2)
\end{equation}
where $A_V$ is the visual extinction and $N(H_2)$ is equal to 1.6 $\times 10^{21}$ cm$^{-2}$, the hydrogen column density corresponding to $A_V$ = 1 mag.

As seen in Figure \ref{fig:gas_ch3oh_iras}, when comparing the models with the range of column densities of methanol recorded by \cite{holdship2019}, and assuming the same $A_{v}$ in both model and observations, it cannot be established which is the dominant formation pathway since both networks lead to methanol abundances that are within the observed range. 

\section{Impact of H-abstraction route on precursors}\label{breakdown}

The addition of the new route in the chemical network has a significant impact on the immediate methanol precursors, \ce{CH3O} and \ce{H2CO}, as well as the by-product \ce{HCO}, particularly in the ice phase. Since the three molecules are significantly affected by shock scenarios, we will discuss their shock behaviour as well as how the radical-molecule reaction affects their overall chemical pathway over time.

Generally speaking, in both low- and high-velocity shocks as well as default and current networks, \ce{HCO}, \ce{H2CO}, and \ce{CH3O} solid-abundances drop dramatically (below $10^{-15}$) as the shock wave propagates. Similar to methanol, their solid-state abundances increases again once the shock wave dissipates. We shall discuss each precursor in turn.

\subsection{Formyl radical (\ce{HCO})}

The formation of \ce{HCO_{(bulk)}} is influenced by the insertion of the H-abstraction route into the chemical network. In scenarios with or without shock, \ce{HCO_{(bulk)}} formation is attributed to reactions \eqref{eq:form hco 2}\citep{simons2020} and \eqref{eq:form hco 3}\citep{fedoseev2016}, as well as the H-abstraction reaction. The latter is the preferential reaction at $n_{H_{2\:(init)}}=10^{4}\:cm^{-3}$, while in other models, especially at high $n_{H_{2}}$, its importance increases as $T_{k}\:(t)$ increases. 

\begin{align}
&\ce{H_{(bulk)} + CO_{(bulk)} -> HCO_{(bulk)}}, \label{eq:form hco 2}\\
&\ce{NH2_{(bulk)} + H2CO_{(bulk)} -> NH3_{(bulk)} + HCO_{(bulk)}},\label{eq:form hco 3}
\end{align}

\begin{figure}[h]
            \centering
            \resizebox{\columnwidth}{!}{
            \includegraphics[width=350pt]{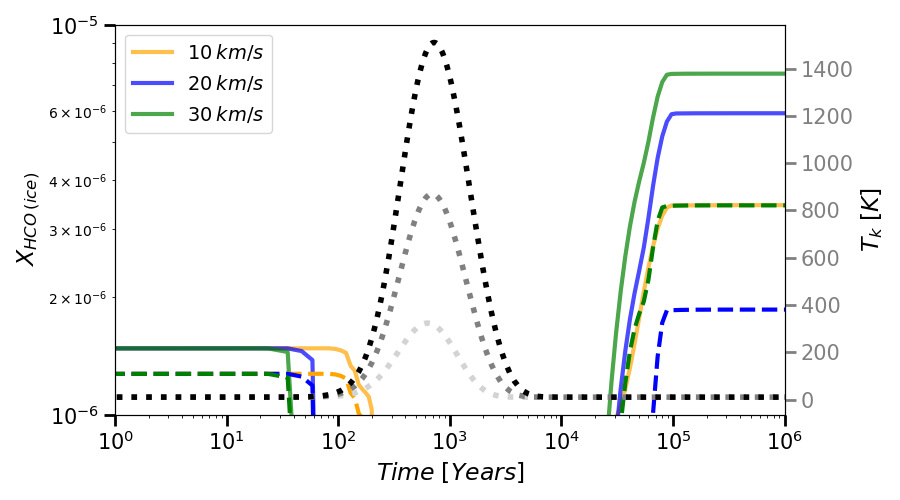}}
            \caption{Chemical abundances of ice-phase \ce{HCO} as a function of time in slow C-shock (10-20 km/s) and moderate-high C-shock (30 km/s) models with a pre-shocked gas density volume of $n_{H_{2\:(init)}}=10^{4}\: cm^{-3}$ and $\zeta=10\zeta_{0}$. The dashed and solid lines correspond to the default and current schemes, respectively. The evolution of gas temperature, $T_{k}\:(t)$, is depicted by the gray dotted line.} \label{fig:grain_hco}
        \end{figure}

With respect to \ce{HCO}, in models with $\zeta=10\zeta_{0}$ and $n_{H_{2\:(init)}}=10^{4}\:cm^{-3}$, $\frac{X_{default}}{X_{current}}\geq2$ is observed during the post-shock epoch of low velocity ($\varv_{shock}=10-20 \:km/s$) and moderate-high velocity C-shocks ($\varv_{shock}=30 \:km/s$). When comparing the abundance of ice formyl radicals between networks (see Fig. \ref{fig:grain_hco}), it is generally observed that the current network  predicts more \ce{HCO} abundance: $0.76\leq\frac{X_{default}}{X_{current}}\leq0.91$, when $X_{\ce{HCO}_{(ice)}}$ exceeds $10^{-7}$. This is because \ce{HCO} is a by-product in the radical-molecule route. The \ce{HCO} radical is an important astrochemical building block due to its ability  to synthesize bio-relevant COMs, such as glycolaldehyde, ethylene glycol, and methyl formate \citep{woods2013,simons2020,li2022, santos2022}. Therefore, this new reaction must be considered in the overall ice-grain chemical network as it influences the abundance of formyl radical and, hence, the distribution of COMs in the ISM.

\subsection{Formaldehyde (\ce{H2CO})}

\begin{figure}[h]
            \centering
            \resizebox{\columnwidth}{!}{
            \includegraphics[width=350pt]{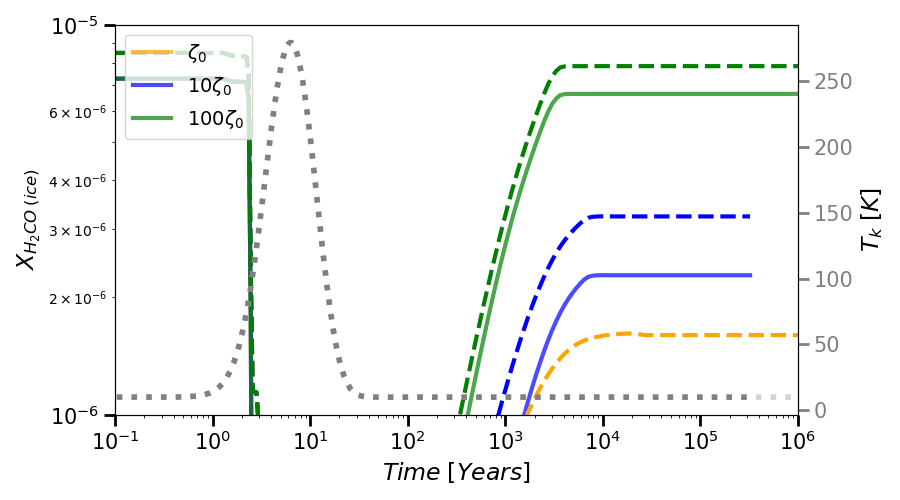}}
            \caption{Chemical abundances of ice-phase \ce{H2CO} as a function of time for slow C-shock models with a pre-shocked gas density volume of $n_{H_{2\:(init)}}=10^{6}\: cm^{-3}$ and $\varv_{shock}=10\:km/s$. The dashed and solid lines correspond to the default and current schemes, respectively. The evolution of gas temperature, $T_{k}\:(t)$, is depicted by the gray dotted line.} \label{fig:grain_formal}
        \end{figure}

In non-shock and shocked scenarios, the H-abstraction reaction dominates the \ce{H2CO} destruction in the grain bulk from $n_{H_{2\:(init)}}=10^{4}\:cm^{-3}$ onwards, becoming more prevalent as $T_{k}\:(t)$ increases. The ratio $\frac{X_{default}}{X_{current}}$ shows an increase with increasing density, reaching values exceeding $2$, particularly in the post-shock epoch of the model with $n_{H_{2\:(init)}}=10^{6}\:cm^{-3}$, $\varv_{shock}=10km/s$, and $\zeta=\zeta_{0}$ conditions, where $\frac{X_{default}}{X_{current}}\geq3.5$ (see Fig. \ref{fig:grain_formal}). It is generally shown that as $n_{H_{2\:(init)}}$ increase in the default scheme, there is more formaldehyde in the solid state compared to the current networks: $1.03\leq\frac{X_{default}}{X_{current}}\leq3.68$, when $X_{\ce{H2CO}_{(ice)}}$ exceeds $10^{-7}$.  This difference is due to the faster radical-molecule rate, which results in more formaldehyde being used to produce ice methanol and hence providing a lower \ce{H2CO} abundance in the current scheme. 

Formaldehyde may serve as a tool for tracing C-type shocks and distinguishing between methanol formation pathways in low-velocity shocked environments. At a shock velocity of 10 km/s, gaseous \ce{H2CO} is enhanced by the shock and traces its full extent.  However, in models with $n_{H_{2\:(init)}}=10^{3}\: cm^{-3}$ and $\zeta=100 \zeta_{0}$, and in all models where $\varv_{shock}$=50 km/s, while it is enhanced initially by the shock, it is then destroyed as the shock increases the gas kinetic temperature.  Interestingly, we can identify C-shock velocities by examining the behaviour of methanol and formaldehyde. Formaldehyde serves as a shock tracer when destroyed at high shock velocities, while methanol is enhanced by the shock and traces its full extent. Therefore, the ratio of these two species may be a good shock diagnostic. 

\begin{figure*}[t]
            \sidecaption
            \includegraphics[width=12cm]
            {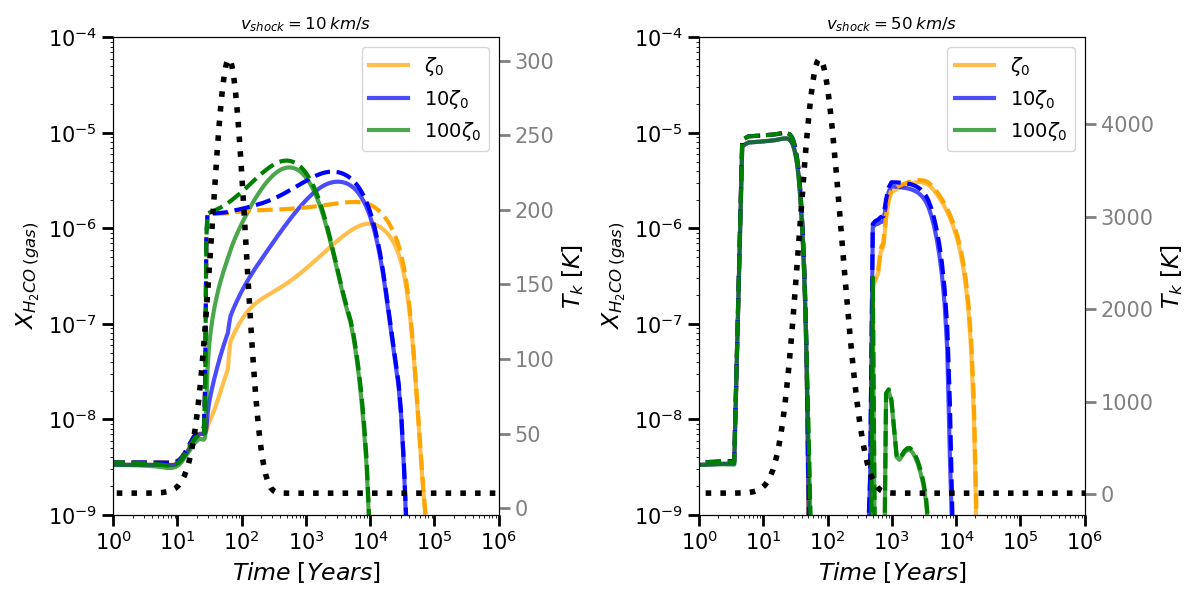}
            \caption{Chemical abundances of gas-phase formaldehyde as a function of time for slow C-shock (left panel) and high C-shock (right panel) models with a pre-shocked gas density volume of $n_{H_{2\:(init)}}=10^{5}\: cm^{-3}$. The dashed and solid lines correspond to the default and current schemes, respectively. The evolution of gas temperature, $T_{k}\:(t)$, is depicted by the gray dotted line.} \label{fig:gas_formal}
        \end{figure*}

In general,  gaseous formaldehyde seems to be a great species to discern between methanol  formation routes. High density ($n_{H_{2\:(init)}}=10^{5}-10^{6}\: cm^{-3}$) and low-velocity C-shock ($\varv_{shock}=10km/s$) models show significant contrast between the default and current networks, with $\frac{X_{default}}{X_{current}}\geq170$ (Fig. \ref{fig:gas_formal}).
In non-shock models, the difference is noticeable as the temperature increases, being higher than $2.5$ as the temperature stabilizes.
In addition, and of critical importance, the observed abundance ratio of \ce{H2CO}/\ce{CH3OH} might distinguish between methanol formation pathways in low-velocity shocked environments. 

\begin{figure}[h]
            \centering
            \resizebox{\columnwidth}{!}{
            \includegraphics[width=345pt]{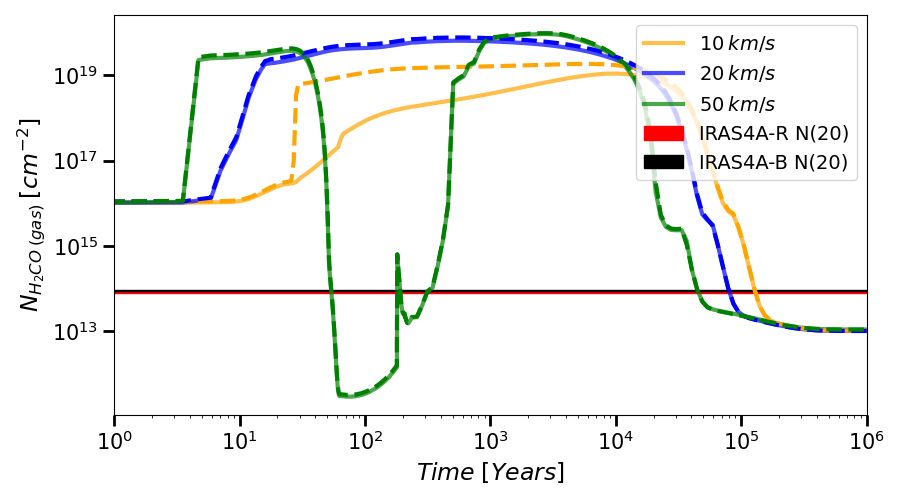}}
            \caption{Column densities of gaseous \ce{H2CO} as a function of time for slow C-shock (10-20 km/s) and high C-shock (50 km/s) models with a pre-shocked gas density volume of $n_{H_{2\:(init)}}=10^{5}\: cm^{-3}$ and $\zeta = \zeta_{0}$. The dashed and solid lines correspond to the default and current schemes, respectively.  The hatches represent the range in \ce{H2CO} column density reported by \cite{gomez2016} for IRAS4A-R, and IRAS4A-B.} \label{fig:gas_h2co_iras}
        \end{figure}

Similarly to our approach with gaseous methanol, we modelled the physical structure traced by \ce{H2CO} in the protostellar outflow sources IRAS4A-R and IRAS4A-B using the initial gas densities ($n_{H_{2}:(init)}$) inferred by \cite{gomez2016}. Figure \ref{fig:gas_h2co_iras} presents a comparison of gas-phase \ce{H2CO} column densities over time from 10–50 km/s models with the lower limits calculated assuming a gas temperature of 20 K, as reported by \cite{gomez2016}, for both IRAS4A-R and IRAS4A-B. As can be seen, it remains unclear which formation pathway is dominant, as both networks yield formaldehyde abundances that fall within the observed range.

\subsection{Methoxyl radical (\ce{CH3O})}

\begin{figure}[h]
            \centering
            \resizebox{\columnwidth}{!}{
            \includegraphics[width=350pt]{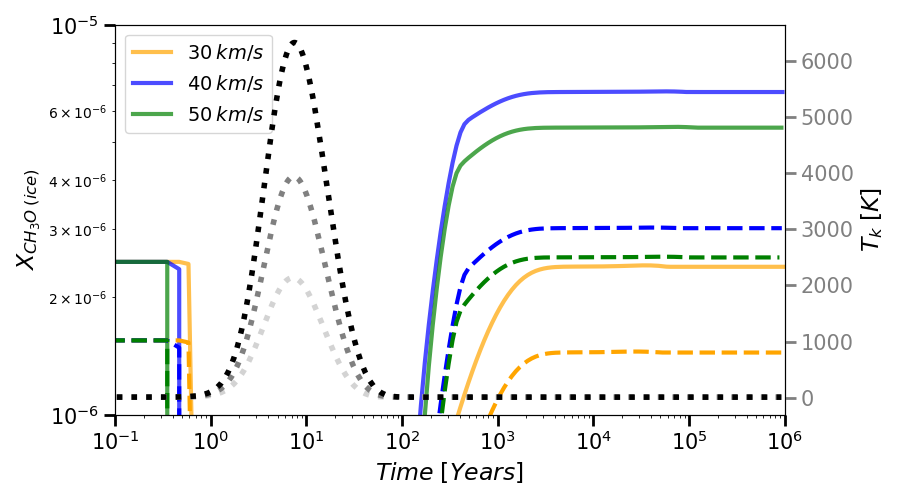}}
            \caption{Chemical abundances of ice-phase \ce{CH3O} as a function of time for C-shock models with a pre-shocked gas density volume of $n_{H_{2\:(init)}}=10^{6}\: cm^{-3}$ and $\zeta=\zeta_{0}$. The dashed and solid lines correspond to the default and current schemes, respectively. The evolution of gas temperature, $T_{k}\:(t)$, is depicted by the gray dotted lines.} \label{fig:grain_methoxi}
        \end{figure}

From $n_{H_{2\:(init)}}=10^{4}\:cm^{-3}$ onwards, the radical-molecule route (Rxn.\eqref{eq:abstract bulk}) and layer transfer are the primary destruction reactions for \ce{CH3O_{bulk}}. For most time steps, the current network produces in general more methoxyl radical, due to the change in branching ratios.  We observed that as the $\varv_{shock}$ increases, the \ce{CH3O_{(bulk)}} abundance  is greater than when using  the default network, with a $\frac{X_{default}}{X_{current}}\geq2$ (see Fig. \ref{fig:grain_methoxi}).
       
Finally, the biggest ratios are observed at grain formation following the cessation of shock influence. It can be reasonably assumed that enhancing the resolution will enable us to distinguish the preferred route of methanol formation throughout the formation and growth of the grain, particularly in its precursors, as it enables us to detect smaller fluctuations and gain deeper insights into the processes that drive molecular evolution in space. For instance, it is anticipated that by improving the resolution of the observations, the $\sim10\%$ increase in \ce{HCO_{(ice)}} seen in the models at $n_{H_{2\:(init)}}=10^{4}\:cm^{-3}$ could be contrasted.

\section{Conclusions}

We theoretically investigate the impact of the radical-molecule H-atom abstraction route in methanol formation under shock-influenced environments. We confirmed that, with  the chemical model \texttt{UCLCHEM}, the radical-molecule pathway is the most favoured reaction for the final step in the overall \ce{CO} hydrogenation mechanism leading to \ce{CH3OH} formation in the bulk-phase of interstellar ices. This finding is  consistent with \cite{simons2020} and \cite{santos2022}. However, its dominance is dependent on the gas kinetic temperature $T_{k}$, gas volume density $n_{H_{2}}$, velocity of the C-shock wave $\varv_{shock}$, and cosmic ray ionisation rate $\zeta$.

The consideration of the radical-molecule route or the change in the branching ratios in the chemical network does not affect the ability of \ce{CH3OH} to trace C-shocks. This is because in both schemes, the increase in gaseous \ce{CH3OH}  abundance is achieved by releasing \ce{CH3OH}  from a solid state into the gas phase through sublimation and sputtering.

The incorporation of the radical-molecule H-atom abstraction route into  chemical networks has a substantial impact on the abundance of the methanol precursors, \ce{CH3O} and \ce{H2CO}, as well as on the abundance of the by-product \ce{HCO}. The change in the abundance of these chemical species upon introducing the reaction is more significant than the change observed for \ce{CH3OH}, particularly in the ice phase.

Gaseous formaldehyde may serve as a tool for tracing C-type shocks and be used to discern between methanol formation pathways in low-velocity C-shocked environments, as its abundance undergoes a more significant change than that of \ce{CH3OH} when the H-atom abstraction reaction is included in UCLCHEM's chemical network. Consequently, the abundance ratio $\frac{X_{\ce{H2CO}}}{X_{\ce{CH3OH}}}$ could serve as a useful diagnostic tool for distinguishing between methanol formation pathways in these environments.

\begin{acknowledgements}
We acknowledge the very helpful comments from an anonymous referee which led to an improvement of this manuscript. This work is financially supported by advanced ERC funding (grant ID: 833460, PI: Serena Viti) within the framework of MOlecules as Probes of the Physics of EXternal (MOPPEX) project. EMR gratefully acknowledges the Leiden/ESA Astrophysics Program for Summer Students (LEAPS) 2022, organised by Leiden University and the European Space Agency, and the support provided by the Erasmus+ ICM Grant and Observatorio Astronómico Nacional de Colombia. The project leading to this publication was initiated within the framework of this programme.
\end{acknowledgements}

\bibliographystyle{aa} 
\bibliography{aa56145-25.bib}
\end{document}